# Coupled Decorated Membrane Resonators with Large Willis Coupling

Joshua Lau, Suet To Tang, Min Yang, and Z. Yang[*]

Department of Physics, Hong Kong University of Science and Technology, Clear Water Bay, Kowloon, Hong Kong, China

**Abstract**

We report the experimental studies on acoustic scatterers consisting of a pair of coupled decorated membrane resonators (DMR's) that exhibits near extreme contrast in reflection asymmetry and strong Willis coupling coefficient with amplitude of 2.5 and polarizability cross term of $1.1 \times 10^{-8}$ ms$^2$. The design is based on the theoretical analysis using the Green's function formulism we developed earlier. The clean line shapes of the real and the imaginary parts of the Willis coefficient clear show, even by visual inspection, that the Kramers-Kronig relations are satisfied. Theoretical analysis agrees well with the major features of the experimental results.

I    INTRODUCTION

The Willis materials first proposed by J. R. Willis in 1981 [1] are a new class of bianisotropic meta-materials that exhibit strain-velocity coupling in elasticity or pressure-velocity coupling in acoustics. Further theoretical studies by Milton *et al.* enriched the conceptual contents with simple yet clear model systems [2, 3]. Norris *et al.* developed an analytical homogenization formulism for three dimensional periodic elastic systems [4] which could provide some guidance for experimental realization of the meta-materials. Nassar *et al.* revisited the original theory by Willis [5], and derived a one dimensional discrete periodic system. Muhlestein *et al.* further analyzed the reciprocity, passivity and causality in Willis materials [6], and predicted that the Willis cross coupling coefficient must have the line shapes that follow the Kramers-Kronig relation dictated by causality. Sieck *et al.* further developed a source driven multiple scattering scheme [7] that emphasizes the physical origin, in particular the asymmetry of the building blocks of the media, and pointed out the necessity of including the Willis coupling in retrieving the macroscopic dynamic parameters of inhomogeneous media that are consistent with reciprocity, passivity and causality. Quan *et al.* derived general bounds on the response of acoustic scatterers including the Willis coupling coefficient based on energy conservation [8]. Li *et al.* reported systematic design and successfully demonstrated wavefront manipulation by bianisotropic metasurface [9]. However, as the reflection asymmetry was only in the phase spectra but not in the amplitude, the obtained cross coupling polarizability was purely imaginary [10] rather than complex. Su *et al* derived several simple formulae to retrieve the polarizability tensor for subwavelength acoustic scatterers using a finite set of scattering amplitudes [10]. They have also designed and numerically investigated a type of scatterers based on asymmetric Helmholtz resonators, and found the maximum cross coupling coefficient at resonance to be about $10^{-9}$ ms$^2$.

In the area of experimental investigations, Koo *et. al* used meta-atoms made by elastic membranes mounted on rectangular rigid frames to reach the whole constitutive parameter space, including the Willis coupling term [11]. However, the effect was small because the meta-atoms were not designed for optimum Willis meta-materials. Muhlestein *et al.* investigated a structure with an elastic membrane and a perforated paper separated by a spacer [12]. However, like in the case of Ref. 10, the reflection asymmetry occurred only in the phase spectrum but not in the amplitude, leading to the real part of the coupling coefficient close to zero, and failed to rise as a resonance predicted by theory. Liu *et. al* investigated a structured beam with asymmetric unit cells [13]. Reflection asymmetry in both the amplitude and the phase spectra was observed, but from the obtained line shape of the Willis coupling coefficient it is hard to tell whether the Kramers-Kronig relations are satisfied, because only a portion of an apparent resonance feature was experimentally observed. Ma *et. al* studied asymmetric bilayer structures and found the maximum Willis coefficient to be around 0.5 [14]. However, in the frequency range where the Willis coupling is relatively large, the transmission is always very small, which means that the scatterer is almost opaque. To date, the observed experimental effects of Willis coupling are mostly in the region of weak perturbation [8], so there is no clear experimental evidence of how large the coupling can be, and whether the Willis coupling term satisfies the Kramers-Kronig relations, even though it is demanded by causality.

In this paper, we first present the formulism for the Willi's coupling in the framework of local resonators homogenization scheme that focuses on the resonant modes of the acoustic scatterers [15, 16], and point out the general strategy in the design of such scatterers with large Willi's coupling. We then report the experimental studies on an acoustic scatterer consisting of a pair of coupled decorated membrane resonators (DMR's) that exhibits near extreme contrast in reflection asymmetry and strong Willis coupling coefficient with amplitude as high as 2.5. The value is $1.1 \times 10^{-8}$ ms$^2$ in terms of the cross term in the polarizability tensor $\alpha^{pv'}$. The design is based on the theoretical analysis using the Green's function formulism we developed earlier [15]. The clean line shapes of the real and the imaginary parts of the Willis coefficient clear show, even by visual inspection, that the Kramers-Kronig relations are satisfied. In the frequency regions where there are strong coupling among the resonant modes of the scatterer, large Willis coupling can be obtained while the transmission of the scatterer is still fairly large (~ 20 %).

## II    THEORY

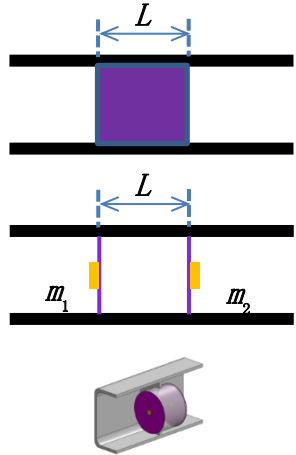

Figure 1    Schematics of an elastic body (Top) and a double-DMR scatterer (Middle) in a waveguide. The bottom one shows the pictorial view of the sample.

Consider an elastic body in an acoustic waveguide as shown in the top of Fig. 1. The length of the object measured from its two end surfaces is $L$. The positions of the two surfaces are at $x = \pm a$, respectively, where $a \equiv \frac{L}{2}$. For acoustic waves with wavelength much larger than the cross section dimension of the waveguide, the surface responses of the elastic body, i. e., its displacement at $x$ on the surface under a surface excitation at $x'$, can be described by the following expression using the Green's function formulism [15, 16]:

$$G(x, x') = \sum_n \frac{u_n(x) u_n^*(x')}{\rho_n (\omega_n^2 - \omega^2 - i\omega\beta_n)} \qquad (1),$$

where $u_n(x)$ is the vibration field of the $n$-th eigenmode, $\omega_n$ is the angular frequency of the $n$-th eigenmode, $\beta_n$ is the dissipation, $\rho_n \equiv \int \rho(\vec{r}) u_n^2(\vec{r}) dv$, and $\rho(\vec{r})$ is the mass density distribution of the elastic body. The uniform stress $\sigma$ excitation can be expressed as the boundary load $\sigma E_-(x')$, and the momentum density $P$ excitation can be expressed as $i\omega P L E_+(x')$, where

$$E_\pm(x') = \delta(x'+a) \pm \delta(x'-a) \qquad (2).$$

Define

$$u_\pm(\pm a) \equiv \iint G(x, x') E_\pm(x') \delta(x \mp a) dx dx' = \sum_n \frac{(u_n(a) \pm u_n(-a)) u_n(\pm a)}{\rho_n(\omega_n^2 - \omega^2 - i\omega\beta_n)} \qquad (3),$$

The strain and the velocity induced by a stress load $2\sigma E_-(x')$ are respectively

$$\epsilon_\sigma = \frac{u_-(a) - u_-(-a)}{L} = \sigma \sum_n \frac{2(u_n(a) - u_n(-a))^2}{L\rho_n(\omega_n^2 - \omega^2 - i\omega\beta_n)} \equiv \frac{2G_{11}\sigma}{L} \qquad (4A),$$

and

$$v_\sigma = i\omega \frac{u_-(a) + u_-(-a)}{2} = i\omega\sigma \sum_n \frac{u_n^2(a) - u_n^2(-a)}{\rho_n(\omega_n^2 - \omega^2 - i\omega\beta_n)} \equiv i\omega G_{01}\sigma \qquad (4B).$$

Likewise, the strain and the velocity induced by a momentum load $i\omega P L E_+(x')$ are respectively given by

$$\epsilon_P = \frac{u_+(a) - u_+(-a)}{L} = \sum_n \frac{u_n^2(a) - u_n^2(-a)}{\rho_n(\omega_n^2 - \omega^2 - i\omega\beta_n)} i\omega P \equiv i\omega G_{01} P \qquad (4C),$$

and

$$v_P = i\omega \frac{u_+(a) + u_+(-a)}{2} i\omega P L = i\omega \sum_n \frac{(u_n(a) + u_n(-a))^2}{2\rho_n(\omega_n^2 - \omega^2 - i\omega\beta_n)} \equiv \frac{-\omega^2 L P G_{00}}{2} \qquad (4D).$$

The total strain and the total velocity due to both stimuli are

$$\begin{pmatrix} \epsilon \\ v \end{pmatrix} = \begin{pmatrix} \epsilon_\sigma + \epsilon_p \\ v_\sigma + v_P \end{pmatrix} = \begin{pmatrix} \dfrac{2G_{11}}{L} & i\omega G_{01} \\ i\omega G_{01} & -\dfrac{\omega^2 L G_{00}}{2} \end{pmatrix} \begin{pmatrix} \sigma \\ P \end{pmatrix} \quad (5).$$

Compare with the constituent equation $\begin{pmatrix} \sigma \\ P \end{pmatrix} = \begin{pmatrix} C & D \\ D & \rho \end{pmatrix} \begin{pmatrix} \epsilon \\ v \end{pmatrix}$, we have

$$C = \frac{LG_{00}}{2\Pi} \quad (6A),$$

$$\rho = \frac{-2G_{11}}{\omega^2 L \Pi} \quad (6B),$$

$$D = \frac{-iG_{01}}{\omega \Pi} \quad (6C),$$

where $\Pi \equiv G_{11}G_{00} - G_{01}^2$. For pure dipolar or monopolar resonances $G_{01} = 0$, and Eqs. (6A) and (6B) reduce to the form derived previously [14].

Compare to Ref. 12, the Willis coefficient is $\psi = \dfrac{D}{i\omega}$, and the asymmetry coefficient is $W = \dfrac{-iD}{\sqrt{C\rho}}$. To realize large Willis coefficient, $G_{01}^2$ must be comparable in strength with $G_{00}G_{11}$. One such possibility, according to Eq. (4B), is for $u_n(a)$ to be much larger than $u_n(-a)$, i. e., the displacements of the two surfaces of the elastic body are highly asymmetric.

In terms of experimental investigations, as the Willis coeffocient is given by [12]

$$W = \frac{r - r_B}{i\sqrt{(1 - rr_B + t^2)^2 - 4t^2}} \quad (7A),$$

and in terms of the cross term in the polarizability tensor [10],

$$\alpha^{pv'} = \frac{S_{WG}}{i\omega c}(r - r_B) \quad (7B),$$

where $S_{WG}$ is the cross section area of the waveguide, which in this case is 0.01 m², $c$ is the sound speed in air, $r$ and $r_B$ are the reflection coefficients in the two opposite incident directions, and $t$ is the transmission. The coefficient would be large when one of the reflection coefficient is small and the other is close to 1. Such requirements make asymmetric metasurfaces [17] attractive candidates with large Willis coefficient.

## III EXPERIMENTAL RESULTS AND ANALYSIS

Decorated membrane resonators (DMR's) have been shown to have a number of extraordinary properties [15, 17 – 21]. In this work we use a double-DMR structure with two DMR's separated by a 40 mm sealed air column in between, as schematically shown in the middle and bottom parts of Fig. 1, as a device to realize large Willis coefficient. The radius of the platelet in DMR-1 is 4.8 mm, while that in DMR-2 is 4.0 mm. The diameter of the membrane in both DMR's is 40 mm, and the thickness is 0.02 mm. The remaining cross section area of the waveguide was blocked by a hard plastic plate with an opening to mount the sample. Some plasticine was added to the steel platelets to fine tune the properties of the device. Define asymmetry scale from AS-0 to AS-4, starting from AS-0, the mass of the platelet on DMR-1 is $m_1 = 63.7$ mg and that on DMR-2 is $m_2 = 57.7$ mg (fixed). The mass of the platelet on DMR-1 was then systematically reduced by removing the plasticine on the platelet to $m_1 = 62.2$ mg to reach AS-1, $m_1 = 58.0$ mg to reach AS-2, $m_1 = 51.7$ mg to reach AS-3, and finally $m_1 = 44.0$ mg to reach AS-4. All the transmission and the reflection measurements were done on an impedance tube system reported in our earlier works [21]. At each AS-n setting both the reflection spectra for waves incident to the DMR-1 side (front) and to the DMR-2 side (back) were measured, along with the transmission spectra.

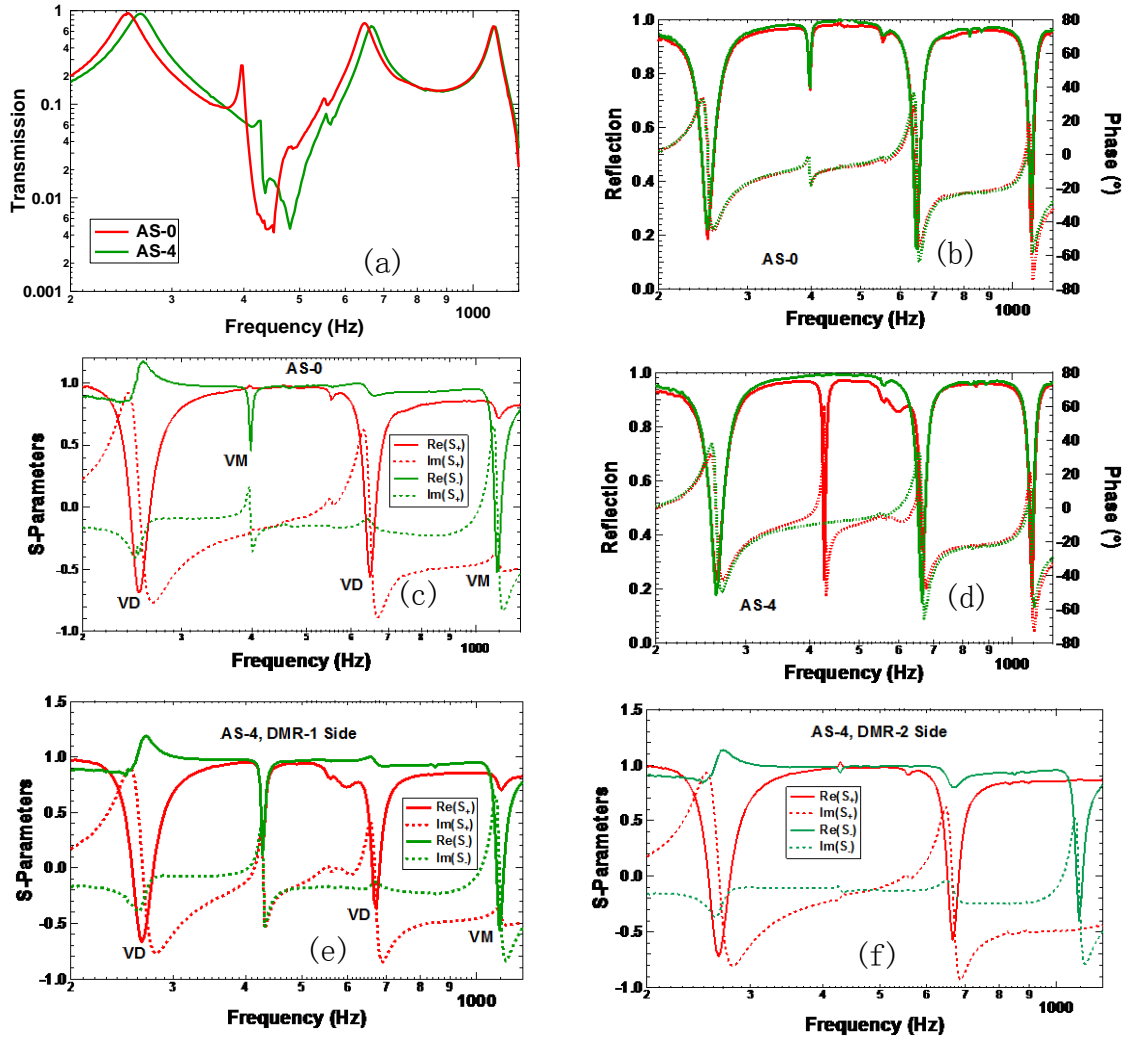

Figure 2    (a) The experimental transmission spectra of the double-DMR structure in its least asymmetric form (AS-0) and the most asymmetric form (AS-4); (b) The DMR-1 side reflection (red curves) and the DMR-2 side reflection (green curves) spectra of AS-0. The solid curves are amplitudes and the dashed curves are phase spectra; (c) The corresponding S-parameters for AS-0. The solid curves are the real parts, and the dashed ones are the imaginary parts; (d) The corresponding reflection spectra for AS-4; (e) The S-parameters for AS-4 with the reflection spectrum from the DMR-1 side; (f) The S-parameters for AS-4 from the DMR-2 side.

Figure 2(a) is the transmission spectra of the two extreme cases AS-0 and AS-4. The spectrum for AS-0 exhibits three major transmission maxima near 250 Hz, 650 Hz, and 1090 Hz, one minor maximum at 396 Hz, and two transmission minima near 440 Hz and 1214 Hz. The DMR-1 side (red curve) and the DMR-2 side (green curve) reflection spectra are shown in Fig. 2(b), which are nearly the same. The reflection minima match the transmission maxima in Fig. 2(a). To examine the nature and the symmetry of the resonant features in the reflection and the transmission spectra, we plot the scattering parameters (S-parameters) $S_\pm \equiv r \mp t$ in Fig. 2(c). Since there is little reflection asymmetry for AS-0, only the S-parameters for the reflection on the DMR-1 side are shown. We first look at the resonant feature of the $S_-$ spectrum (the green curves), which exhibits a resonant feature at the same frequency of the reflection minimum and the minor transmission maximum. The $S_+$ spectrum, on the other hand, remains to be nearly unity. This is a typical feature of a velocity monopole (VM) resonance [22], as $1 = S_+ \equiv r - t$, or $1 - r = -t$, implies that the velocity field near the sample is anti-symmetric. For the same reason, it can be determined that the resonance near 1100 Hz is also VM. For the two major transmission maxima and reflection minima near 250 Hz and 650 Hz, the $S_+$ spectrum exhibits strong resonances, while $1 \approx S_- = r + t$, or $1 - r = t$, implying that the velocity field is symmetric. These two spectral features are therefore due to the velocity dipolar (VD) resonances of the sample. The transmission minimum near 440 Hz is due to the anti-resonance of the two VD adjacent resonances, while the minimum near 1200 Hz is probably due to the anti-resonance of the two adjacent VM resonances, as the anti-resonances due to monopolar resonances are not yet well studied due to the interference of the strong dipolar resonances which are usually nearby.

It is known that the perfect absorption hybrid resonance of either one of the two DMR's could occur at the frequency a bit lower than the anti-resonance [17, 19] if the other is or acts as a hard wall. This can be realized by increasing the hybrid resonance frequency of, say, DMR-1 until it matches the anti-resonance of DMR-2. While in our previous work we used external voltage to tune the eigenmodes of one DMR [17], in this work it was done by decreasing the platelet mass of DMR-1. Near perfect absorption of the DMR-1 side of the sample can be realized when the hybrid resonance frequency of the DMR-1 matches the anti-resonance of DMR-2, which acts as a hard wall and the air column between the two DMR's acts like a cavity for the hybrid resonance [19].

The transmission spectrum (green curve) for AS-4 in Fig. 2(a) does not show much distinct differences from the AS-0 one, except that the two dipolar transmission maxima have blued shifted by a small amount. The two reflection spectra, however, are distinctly different near 428 Hz, where the DMR-2 side reflection is near unity while the DMR-1 side reflection is 0.23, leading to the DMR-1 side absorption being over 95 % while the DMR-2 side absorption being near zero. The sample is now in a highly asymmetric form near 428 Hz.

Shown in Fig. 2(e) and 2(f) are the DMR-1 side and the DMR-2 side S-parameter spectra. While other resonance features remain almost the same as in AS-0, the feature at the maximum reflection asymmetry frequency of 428 Hz is distinctly different from the VM near 400 Hz in Fig. 2(c). Here both the $S_+$ and the $S_-$ spectra exhibit a resonant feature, indicating that when the sound waves are incident on the DMR-1 side, they excite a resonance which possess neither dipolar nor monopolar symmetry. Or one may say they have simultaneously excited a dipole and a monopole of equal strength. For sound waves incident on the DMR-2 side, however, there is almost no resonance feature around 428 Hz. The sound waves just 'see' a hard wall. Therefore, near 428 Hz the properties of the sample is highly asymmetric, a condition favorable for large Willis coupling.

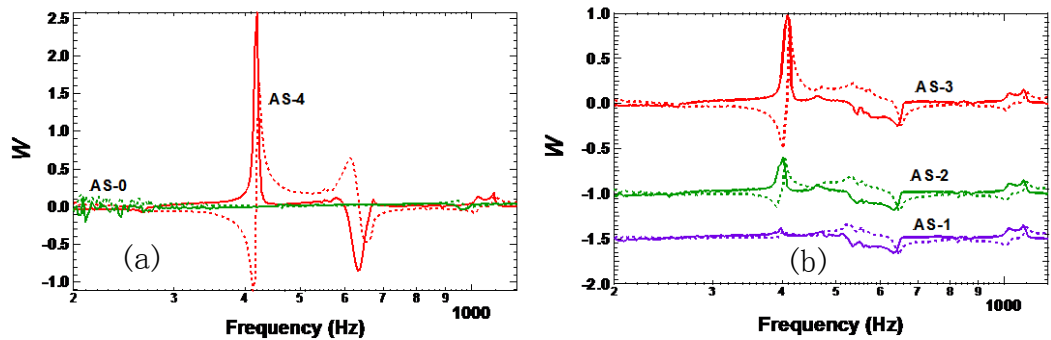

Figure 3 (a) The Willis parameter spectra for AS-0 (green curves) and AS-4 (red curves); and (b) The Willis parameter spectra for AS-1 through AS-3. The solid curves are for the real part and the dashed curves are for the imaginary parts. For clarity, the curves for AS-2 are down shifted by 1.0, and that for AS-1 are down shifted by 1.5.

The Willis parameter $W$ for the two extreme cases (AS-0 and AS-4) obtained by using Eq. 7A are shown in Fig. 3(a), along with the ones for the three intermittent cases, namely AS-1 through AS-3 in Fig. 3(b). A major feature with complete resonant line shape near 428 Hz (W-1) is clearly seen. For the extreme case of AS-4, the real part of the W-1 Willis parameter is over 2.5, which is as large as 20 times the reported one so far [12]. In terms of the cross term in the polarizability tensor $\alpha^{pv'}$, the value is $1.1\times10^{-8}$ ms$^2$, which is 10 times higher than the value found in the device designed and analysed in Ref. 10. Here the Willis effect is no longer a perturbation, but a strong resonance-like feature. The amplitudes of the Willis parameter decrease as the sample becomes less asymmetric, and finally reaches near zero for the case of AS-0.

A second resonance feature (W-2) in the AS-4 Willis parameter spectrum is also present near 620 Hz. Its sign is opposite of W-1, and its peak value is 0.8, still much larger than the one reported so far [12]. The S-parameter spectra shown in Fig. 2(e) and 2(f) indicate that it is neither a monopole nor a dipole. It is likely the hybrid of the highly asymmetric VM near 428 Hz and the strong VD resonance near 650 Hz. As these resonances are close by in frequency, at any frequency the vibration of the sample is usually a mixture of these nearby resonances. For practical reasons to use several Willis units in series, W-2 has the advantage over W-1 in that for W-1, the transmission in the W-1 frequency range is very small (~ 2%), as one surface (DMR-2) acts like a hard wall. For W-2, the transmission in the spectral range is about 10 times higher (~ 20 %).

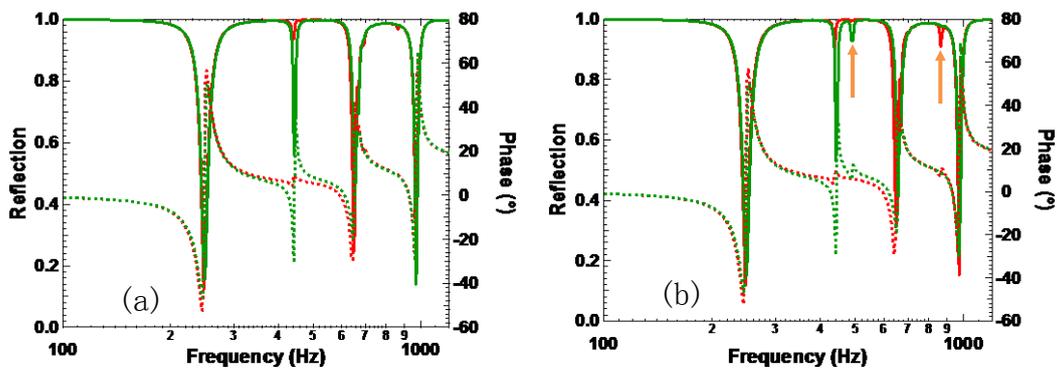

Figure 4 (a) Asymmetric reflection from simulations when both platelets of the DMR's are at the membrane center; (b) The platelet of DMR-1 is 1 mm off center.

Our theoretical investigations show that the W-2 feature is due to the slight eccentricity of the platelet position which was accidentally introduced in the sample fabrication. Figure 4(a) shows the DMR-1 side and the DMR-2 side reflection spectra obtained by simulations for AS-4 with the platelets of both DMR's precisely at the centers. The materials parameters and the pre-tension used in the simulations are the same as in Ref. 20. The simulation results can reproduce all the major features of the sample observed in experiments, except for the asymmetric feature W-2. Vibration pattern analysis shows that the reflection dips around 250 Hz and 650 Hz are due to dipolar resonances, which is consistent with the experimental ones at almost the same frequencies. The asymmetric reflection dip from simulations is at 442 Hz, which is very close to the experimental 428 Hz. The 2$^{nd}$ VM from simulations is at 1000 Hz, which is off by about 100 Hz from the experimental one. By placing the platelet of DMR-1 off the center by 1.0 mm, the simulation results reproduce semi-quantitatively the W-2 feature, as is shown in Fig. 4(b) by the orange-color arrow near 500 Hz. Its general position is correct, i. e., between the reflection asymmetry and the 2$^{nd}$ VD resonance, even though the exact position is somewhat different from the one observed in the experimental spectrum. The simulations also predict a second weak reflection asymmetry near 900 Hz, which is absent in the experimental spectra. Overall, the preliminary simulations provide a semi-quantitative picture of the underline mechanism of large reflection asymmetry.

One can tell by simple visual inspection that the line shapes of both W-1 and W-2 have all the major features of textbook Lorentzian type response functions that satisfy the Kramers-Kronig relations. For example, the maximum amplitude of the real part coincides with the zero-point of the imaginary part. The maximum and the minimum points of the imaginary part coincide with the half amplitude points of the real part. The asymmetric double-DMR structure studied here has proven itself to be a promising example for devices with large and resonant type Willis coupling.

## IV    SUMMARY

In the quest for large Willis coefficient, one would intuitively use scatterers with highly asymmetric structures. However, in the present work the two DMR's of the sample are only different by a seemingly small margin. The relative difference in the platelet masses is less than 20 %. Large Willis coefficients are realized by the large asymmetry of the acoustic properties of the two surfaces of the scatterers, rather than seemingly large structural asymmetry. The eccentricity of the platelet position off the membrane center could introduce large Willies coefficient in the off-resonant frequency range.

**Acknowledgement**

This work was supported by AoE/P-02/12 from the Research Grant Council of the Hong Kong SAR government.

**References**

[1]   J. R. Willis, *Wave Motion* **3**, 1 (1981).
[2]   G.W. Milton and J. R. Willis, *Proc. R. Soc.* **A 463**, 855 (2007).
[3]   G. W. Milton, *New Journal of Physics* **9**, 359 (2007).
[4]   A. N. Norris, A. L. Shuvalov, and A. A. Kutsenko, *Proc. R. Soc.* **A 468**, 1629 (2012).
[5]   H. Nassar, Q.-C. He, N. Auffray, N. Willis, *J. Mech. Phys. Solids* **77**, 158–178 (2015).
[6]   M. B. Muhlestein, C. F. Sieck, A. Alù, and M. R. Haberman, *Proc. R. Soc.* **A 472**, 20160604 (2016).
[7]   C. F. Sieck, A. Alù, and M. R. Haberman, *Phys. Rev*. **B 96**, 104303 (2017).
[8]   Li Quan, Younes Ra'di, Dimitrios L. Sounas, and Andrea Alù, *Phys. Rev. Lett.* **120**, 254301 (2018).
[9]   J. Li, C. Shen, A. Díaz-Rubio, S. A. Tretyakov, & S. A. Cummer, *Nat. Commun*. 9, 1342 (2018).
[10]  Xiaoshi Su and Andrew N. Norris, *Phys. Rev*. B **98**, 174305 (2018)
[11]  S. Koo, C. Cho, J. Jeong, and N. Park, *Nat. Commun*. **7**, 13012 (2016).

[12] M. B. Muhlestein, C. F. Sieck, P. S. Wilson, and M. R. Haberman, *Nat. Commun*. **8**, 15625 (2017).

[13] Yongquan Liu, Zixian Liang, Jian Zhu, Lingbo Xia, Olivier Mondain-Monval, Thomas Brunet, Andrea Alù, Jensen Li, *arXiv*:1807.02285

[14] Fuyin Ma, Meng Huang, Yicai Xu, and Jiu Hui Wu, Bi-layer plate-type acoustic metamaterials with Willis coupling, *Journal of Applied Physics* **123**, 035104 (2018)

[15] M. Yang, G. Ma, Z. Yang, and P. Sheng, *Phys. Rev. Lett.* **110**, 134301 (2013).

[16] Min Yang, Guancong Ma, Ying Wu, Zhiyu Yang, and Ping Sheng, *Phys. Rev.* **B89**, 064309 (2014)

[17] Songwen Xiao, Suet To Tang, and Z. Yang, *Appl. Phys. Lett*. **111**, 194101 (2017)

[18] Z. Yang, M. Yang, N. H. Chan, and P. Sheng, *Phys. Rev. Lett.* **101**, 204301 (2008)

[19] Guancong Ma, Min Yang, Songwen Xiao, Zhiyu Yang, and Ping Sheng, *Nature Materials* **13**, 873 (2014)

[20] Jun Mei, Guancong Ma, Min Yang, Zhiyu Yang, Weijia Wen & Ping Sheng, *Nat. Commun.* **3**, 756 (2012)

[21] Caixing Fu, Xiaonan Zhang, Min Yang, Songwen Xiao, and Z. Yang, *Appl. Phys. Lett*. **110**, 021901 (2017)

[22] Suet To Tang, Xiaonan Zhang, Chong Meng, and Z. Yang, *Phys. Rev. Appl*. **11**, 014008 (2019)